\documentclass[12pt,]{article}
\usepackage{amsmath}
\usepackage{amsfonts} \usepackage{amssymb}

\numberwithin{equation}{section}

\begin{document}      

\title{Cosmologies with positive $\lambda$: Hierarchies of future
behaviour.}

\author {Helmut Friedrich\\
Max-Planck-Institut f\"ur Gravitationsphysik\\
Am M\"uhlenberg 1\\ 14476 Golm, Germany }

\maketitle

%\maketitle                
 
%\maketitle                

\begin{abstract}

\noindent
Smooth Cauchy data on $\mathbb{S}^3$ for the 
Einstein-$\lambda$-vacuum field equations with cosmological constant $\lambda  > 0$
that are sufficiently close to de Sitter data develop into a solution that admits a smooth conformal boundary ${\cal J}^+$ in its future.
The {\it conformal Einstein equations} determine  
a smooth conformal extension across ${\cal J}^+$ that defines on `the other side'  again a $\lambda$-vacuum solution.
In this article we discuss
to what extent these properties generalize to the {\it future asymptotic behaviour of solutions to the Einstein-$\lambda$ equations with matter}. 
We study FLRW solutions and the Einstein-$\lambda$ equations coupled to 
conformally covariant matter transport equations, to conformally privileged
matter equations, and to conformally non-covariant  matter equations. 
We present recent  results on the
Einstein-$\lambda$-perfect-fluid equations with a 
non-linear {\it asymptotic dust} or  {\it asymptotic radiation} equation of state.

\end{abstract}

% \newpage

{\footnotesize

%\tableofcontents

\vspace{1cm}

\section{Introduction}

Roger Penrose suggested to discuss the asymptotic behaviour of  space-times in terms of extensions of their conformal structure
\cite{penrose:1963}, \cite{penrose:1965}. 
The idea is that a {\it physical space-time} 
$(\hat{M}, \hat{g}_{\mu\nu})$ may admit a smooth extension 
$(M, g_{\mu\nu}, \Omega)$ where $M$ is a smooth manifold with boundary ${\cal J}$,  $g_{\mu\nu}$ a smooth Lorentz metric and  
$\Omega$ a smooth function on $M$, 
so that $M = \hat{M} \cup {\cal J}$, $\Omega > 0$ and 
$g_{\mu\nu} = \Omega^2\,\hat{g}_{\mu\nu}$ on $\hat{M}$, while
$\Omega = 0$ on the set ${\cal J}$ (referred to as the {\it conformal boundary} or short as {\it Scri}).
It may not be easy, or not possible at all,  to find for a given space-time $(\hat{M}, \hat{g}_{\mu\nu})$ 
a smooth conformal extension 
$(M, g_{\mu\nu}, \Omega)$ as above, but if it can be done,
the construction provides precise and complete information about the asymptotic behaviour of the space-time in the neighbourhood of Scri.
 
In the first 20 years following the introduction of the concept there became available related general observations, various discussions of  fields and physical quantities on and near conformal boundaries,
and detailed verifications of the existence of conformal boundaries in the case of a number of important exact solutions with symmetries.
After becoming acquainted  with the idea I tried to understand to what extent the concept applied to general solutions of Einstein's equation 
\begin{equation}
\label{einst}
R_{\mu \nu} [\hat{g}]
- \frac{1}{2}\,R[\hat{g}]\,\hat{g}_{\mu \nu} 
+ \lambda\,\hat{g}_{\mu \nu} = \hat{T}_{\mu \nu},
\end{equation}
with cosmological constant $\lambda > 0$ and energy momentum tensor $ \hat{T}_{\mu \nu}$.

The conformal extension idea will be illustrated here
by a discussion of FLRW models because some of their features will be important for us in the following. By these we understand  
 space-times
with manifold $\mathbb{R} \times S$ and metric 
$\hat{g} =  - dt^2 + a^2\,\hat{h}$, where $a = a(t)$ is a scalar, 
$S$ is a 3-dimensional manifold,  and $\hat{h}$ a
$t$-independent Riemannian metric of constant curvature on $S$
with Ricci scalar $R[\hat{h}] = const. \ge 0$.
The space-times are required to satisfy the Einstein-$\lambda$-perfect-fluid equations with flow field $U = \partial_t$, total energy density $\hat{\rho}(t)$, pressure $ \hat{p}(t)$, and an equation of state
$\hat{p} = w(\hat{\rho})$.

To discuss the solutions in terms of a conformal representation we use the rescalings 
\[
\hat{g}_{\mu\nu} \rightarrow g_{\mu\nu} 
= \Omega^2\,\hat{g}_{\mu\nu}, \quad \quad 
\hat{U}^{\mu} \rightarrow U^{\mu} =   \Omega^{-1}\,\hat{U}^{\mu} , \quad \quad 
\hat{\rho} \rightarrow \rho =  \Omega^{- e}\,\hat{\rho}, \quad
\]
with a conformal factor $\Omega = \Omega(t)$ 
 whose evolution is fixed by the requirements 
that the Ricci scalars of $g$ and $\hat{h}$ satisfy
$R[g] = R[\hat{h}]$ throughout the solution and
$a\,\Omega = 1$, $\,d(a\,\Omega)/dt = 0$ on a given slice 
$\{t = t_*\}$. The constant $e$ will be determined later.

Then $\Omega = a^{-1}$ and in terms of the coordinate $\tau(t) = \tau_* + \int^t_{t_*}\,a^{-1}\,dt$ follows 
\[
g = - d\tau^2 + \hat{h}.
\]
With a linear equation of state 
$w(\hat{\rho}) = w_*\,\hat{\rho}$ where $0 \le  w_* = const. \le 1/3$
the conformal analogues of the well known Friedmann and energy conservation equations, that represent the content of (\ref{einst}),
 then read with the dot denoting 
$d/d\tau $,  
\[
(\dot{\Omega})^2 = \frac{\lambda}{3}
- \frac{R[\hat{h}]}{6}\,\Omega^2 + 
\Omega^{e}\,\frac{\rho}{3}, \quad \quad
\dot{\rho} = \Omega^{-1}\,(3 + 3\,w_* - e)\,\rho\,\dot{\Omega}.
\]
With initial data $\Omega = \Omega_*  = \Omega(\tau_*) > 0$ and
$\rho = \rho_* = \rho(\tau_*) \ge 0$ at $\tau = \tau_*$ this implies
\[
\rho = \rho_* \left(\frac{\Omega}{\Omega_*}\right)^{3 + 3\,w_* - e},
\]
and thus, {\it independent of the choice of $e$}, the conformal Friedmann equation in the form
\[
(\dot{\Omega})^2 = \frac{\lambda}{3}
- \frac{R[\hat{h}]}{6}\,\Omega^2 + 
\frac{\rho_*}{3}\left(\frac{\Omega}{\Omega_*}\right)^{3 + 3\,w_*}.
\]
Let $\Omega_*$ be small (that is $a(t_*)$ be large) so that the right hand side is positive for $\Omega \le \Omega_*$ and choose the sign of the square root and the parameter $\tau$ so that $\dot{\Omega}$ is decreasing while 
$\tau$ is increasing. The equation can then be integrated until
$\Omega  \rightarrow 0$ at some finite value $\tau_{**}$
of the parameter.
 Since $a \rightarrow \infty$ 
and $t \rightarrow \infty$ as $\Omega \rightarrow 0$, 
the hypersurface Scri = $ \{\Omega = 0\} \sim S$ defines a boundary of the physical space-time that represents  future time-like infinity with respect to the physical metric $\hat{g}_{\mu\nu}  = \Omega^{-2}\,g_{\mu\nu}$
and that is space-like with respect to $g_{\mu\nu}$.

That the zero of the function $\Omega$ is given here by a finite value of the conformal coordinate $\tau$ helps to recognize
subtle matter dependent differences in the asymptotic behaviour of  $a(t)$ as $t \rightarrow \infty$. 
In the vacuum case $\rho_* = 0$ we get with
$t_* = \tau_* = 0$, $\lambda = 3$, $S = \mathbb{S}^3$, and
$\hat{h} = h_{\mathbb{S}^3}$ whence $R[\hat{h}] = 6$, 
 the solution $\Omega = \cos \tau$. Then 
$t = \log \tan (\tau/2  + \pi/4)$ and $a = 1/\cosh t$ which gives  the maximally symmetric, geodesically complete, conformally flat {\it de Sitter solution}
\begin{equation}
\label{de Sitter}
\hat{M} = \mathbb{R} \times \mathbb{S}^3, 
\quad \quad 
\hat{g} = - dt^2 + \cosh^2t\,\hat{h}. 
\end{equation}
It is here not so much important for us that the solution can be given explicitly but that we get precise information on the asymptotic behaviour of $\Omega(\tau)$ near Scri. The solution of the ODE above extends smoothly to Scri and in fact beyond. While the `physical' de Sitter metric is defined for $t \in \mathbb{R}$, which is covered by $\tau$ with
$- \pi/2 < \tau < \pi/2$, the cyclic function $\Omega(\tau)$ and the metric $g$ are defined and smooth for $\tau \in \mathbb{R}$, defining a sequence of (isometric) vacuum solutions which are separated by Scri's.

When $\rho_* > 0$ and $0 \le w_* \le 1/3$ different cases occur.
If $0 < w_* < 1/3$ there arise  smoothness and  extension problems. The solutions do approach the value $\Omega = 0$ but
only a few derivatives of the function 
$(\Omega/\Omega_*)^{3 + 3\,w_*}$  have a finite  limit as 
$\Omega \rightarrow 0$.
Some solutions, such as the one obtained with $w_* = 1/9$ for instance, admit a unique extension into a range where $\Omega < 0$
which is smooth where $\Omega \neq 0$ and drops smoothness as 
$\Omega \rightarrow 0$. Others, like the one obtained for $w_* = 1/6$, do not admit an
extension beyond $\Omega = 0$ as solution to the equation above.

However, in the case of {\it pure dust, where $w_* = 0$}, or in the case of {\it (incoherent) pure radiation, where $w_* = 1/3$}, the solutions extend smoothly to $\Omega = 0$ and beyond.
(The unusual word {\it pure} is added here to distinguish these cases clearly from related ones considered later). 
We set for convenience 
$\Omega_* = 1$, $e = 3 + 3\,w_*$ and
consider the 
 {\it case of pure dust with $e = 3$} and 
 the {\it case of pure radiation with $e = 4$}.
 We have then $\rho = \rho_* = const. > 0$ and get
 the conformal Friedmann equation in the form
\[
 (\dot{\Omega})^2 = \frac{\lambda}{3}
- \frac{R[\hat{h}]}{6}\,\Omega^2 + 
\frac{\rho_*}{3}\,\Omega^{e}.
\]
The global solutions and the solution manifold look in these two cases as follows.

 \vspace{.1cm}
 
 \noindent
 {\it Pure dust solutions}. Depending on the  real roots of the polynomial 
 \begin{equation}
 \label{pure-dust-pol}
 P(\Omega) = \frac{\lambda}{3}
- \frac{R[\hat{h}]}{6}\,\Omega^2 + 
\frac{\rho_*}{3}\,\Omega^{3},
\end{equation}
three cases occur (assuming in the following discussions suitable  parameters $\tau$ and signs of the square root of $P$).

\vspace{.2cm}

\noindent
(i) $R[\hat{h}]^3 > 54\,\lambda\,\rho_*^2$: $P$ has roots
$\Omega_- < 0 < \Omega_1 < \Omega_2$. 
There are  solutions that take values in $[\Omega_2, \infty[$ start from a Big Bang  (where $\Omega \rightarrow \infty$, resp.  
$a \rightarrow 0$ for a finite value of the parameter), decrease, reach  their minimum $\Omega_2$ (i.e. $a$ reaches a maximum), increase again and approach a Big Crunch
where $\Omega \rightarrow \infty$ for a finite value of the parameter.
 
 The solutions that take values in $[\Omega_-, \Omega_1[$ are cyclic, 
 oscillating between $\Omega_1$ and $\Omega_-$ and passing through Scri's on the way.
 These approach the de Sitter solution as $\rho_* \rightarrow 0$.

\vspace{.2cm}

\noindent
(ii) $R[\hat{h}]^3 = 54\,\lambda\,\rho_*^2$: Then 
$P = \left(\Omega - \frac{R[\hat{h}]}{3\,\rho_*}\right)^2
\left(\Omega + \frac{R[\hat{h}]}{18}\right)$
with simple root $\Omega_-  = - \frac{R[\hat{h}]}{18}$ and double root 
$\Omega_+ = \frac{R[\hat{h}]}{3\,\rho_*}$. 
The solutions that take values in the domain 
$]\Omega_+, \infty[$ start from a Big Bang, 
decrease monotonously and approach the value $\Omega_+$ without ever assuming it. Similarly, starting from their minimum value $\Omega_-$ the 
solutions which take values in $[\Omega_- , \Omega_+[$ are increasing, pass Scri's,
and approach asymptotically the value $\Omega_+$
 in both directions. Finally there is the time independent solution 
$\Omega(\tau) = \frac{R[\hat{h}]}{3\,\rho_*}$.

\vspace{.2cm}

 \noindent
iii) $0 \le R[\hat{h}]^3 < 54\,\lambda\,\rho_*^2$: $P$ has one real root $\Omega = \Omega_- < 0$. The solutions start with $\Omega > 0$ from a Big Bang, decrease monotonously, pass through a  Scri for a finite value of the parameter, become negative, assume their minimum $\Omega_-$, increase again,
pass through another Scri and approach a Big Crunch where 
$\Omega \rightarrow \infty$.

This solution admits a smooth cyclic extension in the following sense.
Before the limit $\Omega \rightarrow \infty$ is achieved the function $\omega = \Omega^{-1/2}$ is defined and satisfies
\begin{equation}
\label{stiff-equ}
4\,(\dot{\omega})^2 = \frac{\rho_*}{3}
- \frac{R[\hat{h}]}{6}\,\omega^2 
+ \frac{\lambda}{3}\,\omega^6.
\end{equation}
With a redefinition of the constants
this becomes the conformal Friedmann equation above with $w_* = 1$
({\it stiff matter equation of state}) and 
$\lambda$ and $\rho_*$ interchanged.
The condition on $R[\hat{h}]$ above ensures that the polynomial in $\omega$ on the right hand side is positive everywhere and the equation can be integrated across $\omega = 0$. 
Where $\omega < 0$ the transformation $\Omega = \omega^{-2}$ 
connects to a second copy of the solution above and the process can be repeated. 
If the parameter $\tau$ is chosen so that 
$\Omega(0) = \Omega_-$, whence $\Omega(\tau) = \Omega(- \tau)$,
the solution is given 
in terms of Jacobi's elliptic function $cn(u, k)$ 
\cite{lawden} by 
\begin{equation}
\label{cn-solution}
\Omega(\tau) = \Omega_- + \Sigma\,\frac{1 - cn (u(\tau), k)}
{1 + cn(u(\tau), k)},
\end{equation}
where 
$u(\tau) = (\rho_*\,\Sigma/3)^{1/2}\,\tau$, 
the modulus is
$k = (R[\hat{h}] + 2\,\rho_*\,\Sigma - 4\,\rho_*\,\Omega_-)^{1/2}\,
(8\,\rho_*\,\Sigma)^{- 1/2}$, 
$\Sigma = (3\,\Omega_-^2 - \Omega_-\,R[\hat{h}]/\rho_*)^{1/2}$, 
and the root $\Omega_-$ of $P$, that satisfies 
$\Omega_- < - R[\hat{h}]/6\,\rho_*$, is related to $\lambda$ by
$\lambda - R[\hat{h}]\,\Omega_-^2/2 + \rho_*\,\Omega_-^{3} = 0$.

\vspace{.2cm}
 
 \noindent
 {\it Pure radiation solutions}. Depending on the  real roots of the polynomial 
 \begin{equation}
 \label{pure-dust-pol}
Q(\Omega) = \frac{\lambda}{3}
- \frac{R[\hat{h}]}{6}\,\Omega^2 + 
\frac{\rho_*}{3}\,\Omega^{4},
\end{equation}
the following cases occur.

\vspace{.2cm}

\noindent
(i) $R[\hat{h}]^2 > 16\,\rho_*\,\lambda$: Q has four simple roots
$\Omega_1 < \Omega_2 < 0 < \Omega_3 < \Omega_4$. 
There are solutions that take values in $[\Omega_4 , \infty[$, start from a Big Bang, achieve their minimum $\Omega_4$, increase and end in a Big Crunch. 
There are similar solutions that take values in $]- \infty, \Omega_1]$. 

There are cyclic solutions that take values in $[\Omega_2, \Omega_3]$, oscillate between $\Omega_2$ and $\Omega_3$, and pass through Scri's on the way.
These solutions approach the de Sitter solution as $\rho_* \rightarrow 0$.

\vspace{.2cm}

\noindent
(ii) $R[\hat{h}]^2 = 16\,\rho_*\,\lambda$: Then 
$Q = \frac{\rho_*}{3}\left(\Omega^2 - \frac{R[\hat{h}]}{4\,\rho_*}\right)^2$ with the two double roots
$\Omega_{\pm}  = \pm \sqrt{\frac{R[\hat{h}]}{4\,\rho_*}}$. There is a strictly monotonous solution which takes values
in $]\Omega_-, \Omega_+[$, passes through a Scri, and approaches the values $\Omega_-$ and $\Omega_+$ asymptotically. 
There are two strictly monotonous solutions that take values in 
$]\Omega_+, \infty[$ and in
$]- \infty, \Omega_-[$ respectively. 
The first one approaches at one end the value $\Omega_+$ asymptotically and at the other end a Big Bang. The second solution is similar.
Finally there are the time independent solutions 
$\Omega(\tau) = \pm \sqrt{\frac{R[\hat{h}]}{4\,\rho_*}}$.

\vspace{.2cm}
 
 \noindent
(iii) $0 \le R[\hat{h}]^2 < 16\,\rho_*\,\lambda$: $Q$ has no real root.
The solution takes values in $] - \infty, \infty[$. It starts from a Big Bang, decreases with $\Omega > 0$ monotonously, passes through a  Scri,
decreases further, and reaches a Big Crunch where 
$\Omega \rightarrow - \infty$.

Again, this solution admits a smooth cyclic extension in the following sense. With $\omega = - \Omega^{-1}$ close to the Big Crunch the equation for $\Omega$ gives
 \begin{equation}
 \label{omega-pure-rad-equ}
 (\dot{\omega})^2 = \frac{\lambda}{3}\,\omega^4 
- \frac{R[\hat{h}]}{6}\,\omega^2 + 
\frac{\rho_*}{3},
\end{equation}
which is the original equation with the roles of $\lambda$ and $\rho_*$ swapped.
It can be integrated beyond $\omega = 0$ and with 
$\Omega = \omega^{-1}$ be connected to a second copy of the solution for $\Omega$.

\vspace{.2cm}

If the parameter $\tau$ is chosen so that 
$\Omega(0) = 0$ and $\Omega(\tau)$ is increasing with increasing $\tau$ near $\tau = 0$,
the solution is given 
in terms of Jacobi's elliptic functions by 
\begin{equation}
\label{sn-cn-dn-solution}
\Omega(\tau) = \sqrt{f}\,\,\,
\frac{sn(u(\tau), k)}{cn(u(\tau), k)}\,\,
\frac{k' + dn(u(\tau), k)}{1 + dn(u(\tau), k)},
\end{equation}
where $u(\tau) = \sqrt{\rho_*/3}\,(\sqrt{f} + e)\,\tau$,
$f = \sqrt{\lambda/\rho_*}$, 
$e = \sqrt{f/2 + R[\hat{h}]/8\,\rho_*}$,
$k' = \sqrt{1 - k^2}$,
$k = 2\,(e + \sqrt{f})^{-1}\,\sqrt{e\,\sqrt{f}}$.

\vspace{.2cm}

The `fine tuned' cases (ii) will not be of interest to us in the following. 
We shall mainly be interested in the cases (iii) where a Big Bang in the finite past (in terms of physical time) is connected with a 
space-like Scri in the infinite future. These cases admit  the limits $R[\hat{h}] \rightarrow 0$ which are in agreement with the 2018 results of the Planck team. 
We shall later focus on the ends at future time-like infinity where the solutions approach Scri.

\vspace{.1cm}

Of course, in the standard interpretation of GR only a maximal connected space-time with $\Omega > 0$ will be considered as a physical solution. Its conformal extension may just be considered a fun game 
which works because the solutions are
conformally flat and extremely simple.
Heeding Tolman's admonition \cite{tolman:1987}, which reads (with a slight variation) `$\ldots$ we study FLRW models primarily in order to secure definite and relatively simple mathematical problems, rather than to secure a correspondence with known reality $\ldots$' we wonder: Do any of the observations above extend to more general situations ?
We are in particular interested in examples where the fluid flow is not forced to be geodesic and which are not conformally flat
so as to admit gravitational radiation, a concept that FLRW models allow to talk about only in terms of approximations.

Answers of any generality to the question above
need global or semi-global results on suitable Cauchy problems for Einstein's field equations. Moreover, as shown by the subtleties discussed above, they require sharp control  on the asymptotic behaviour of the solutions at least at future time-like infinity. At the time when Penrose put forward his proposal no such results were available. Until the early 1980's the understanding of the Cauchy problem for Einstein's equations was restricted to existence results local in time.

\section{Conformal field equations, stability results.}

Finding initial data which develop into solutions to the Einstein equations that admit smooth conformal boundaries poses subtle problems when the  cosmological constant $\lambda$ vanishes or is negative.
Because the initial slices are not compact there have to be made choices about the fall-off behaviour of the data
 (see  \cite{friedrich:beyond:2015},  \cite{friedrich:pnotp:2018}). 

In the same article where Einstein introduced the cosmological constant, he came to the conclusion that cosmological solutions should be spatially compact \cite{einstein:1917}. In fact, in contrast to the FLRW solutions, which are defined by ODE's, there can be no other choice 
if one whishes to construct geodesically null and time-like future complete solutions to the full Einstein equations. 
There are no natural boundary conditions for the evolution equations if 
$\lambda > 0$.

As indicated above, the Planck team found the Universe to be constrained to be spatially flat to extremely high precision 
\cite {Efstathiou:Gratton:2020}. 
Because no restrictions are given on the `size' of $S$ (in terms of the distances function defined by the prescribed metric $\hat{h}$), we can assume $S$, which we may wish to be simply connected, to be as large as we like so that $R[\hat{h}]$ is lying in the error margin given in \cite {Efstathiou:Gratton:2020} while still being positive.

If the initial slice is compact, which will be assumed in the following, only smoothness and smallness conditions can be imposed on `general' Cauchy data.
The following global non-linear stability result holds \cite{friedrich:1986a}, \cite{friedrich:1986b}.

\vspace{.1cm}

\noindent
{\it On a slice $S = \{t = const.\} \sim \mathbb{S}^3$ of the de Sitter solution
to Einstein's field equations (\ref{einst}) with $\lambda = 3$ and
$\hat{T}_{\mu \nu} = 0$
consider smooth Cauchy data for these equations.  
If these data are (in terms of suitable Sobolev norms) sufficiently close to  the   de Sitter data on  $S$, they develop into solutions to (\ref{einst}) that are time-like and null geodesically complete and admit smooth space-like conformal boundaries at past and future time-like infinity. }

\vspace{.1cm}

Because the cosmological constant can be given any positive value by a conformal rescaling with a constant conformal factor,
the precise value of $\lambda$ is irrelevant here.

The technical basis of this result is a remarkable feature of the Einstein equations. While they are designed to determine a metric, they
can be represented in terms of a conformal factor $\Omega$, the conformal metric $g_{\mu\nu} = \Omega^2\,\hat{g}_{\mu\nu}$ and certain tensor fields derived from them so that they imply with suitable gauge conditions equations that can be hyperbolic even where the conformal factor vanishes or becomes negative, i.e. beyond the domain where $\hat{g}_{\mu\nu}$ is defined
\cite{friedrich:asymp-sym-hyp:1981b}. We refer to these equations
as {\it conformal Einstein equations}. It should be noted that
this name has subsequently also been used for conformal representations of the Einstein equations which were derived for other purposes and do not share the properties used below.

Conformal de Sitter space, given above by
$M = ]- \pi/2,  \pi/2[ \times \mathbb{S}^3$, 
$\,g = - d\tau^2 + \hat{h}$, 
$\,\Omega = \cos \tau$, is a solution to the conformal Einstein equations that extends smoothly, as a solution to the equations and with the same expressions for the metric and conformal factor, beyond the boundaries $\{\tau = \pm \pi/2\}$ to all of
$\mathbb{R} \times \mathbb{S}^3$. Then $\Omega < 0$ on the slices
$\{ \tau = \pm \pi \}$. Consider Cauchy data 
$\bar{g}$, $\bar{\Omega}$, $\ldots$
for the conformal field equations on a slice $\{\tau = \tau_*\}$ with $|\tau_*| < \pi/2$ which are `general' in the sense that symmetries are not necessarily imposed.
If these data are sufficiently close to the conformal de Sitter data induced on 
$\{\tau = \tau_*\}$, general properties of hyperbolic equations 
\cite{kato} guarantee   
that the solution $\bar{g}$, $\bar{\Omega}$, $\dots$ to the conformal field equations which develop from 
the general data also exist on the domain $\{|\tau| \le \pi\}$ and the  conformal factor $\bar{\Omega}$ is negative on $\{\tau = \pm \pi\}$.
The conformal field equations then ensure that there exist two 
hypersurfaces 
${\cal J}^{\pm} \subset \{|\tau| \le \pi\}$
with $\bar{\Omega}|_{{\cal J}^{\pm}} = 0$,  
$d\bar{\Omega}|_{{\cal J}^{\pm}} \neq 0$
that are space-like with respect to $\bar{g}$
and  sandwich a domain 
$\hat{M} \subset \{|\tau| \le \pi\}$ on which $\bar{\Omega}> 0$.
Then $(\hat{M}, \hat{g} = \bar{\Omega}^{-2}\,\bar{g})$ is the desired solution to the Einstein equations.
 
 \vspace{.2cm}
 
The fact that the set of all Cauchy data on $\mathbb{S}^3$ for the Einstein vacuum equations with positive cosmological constant
contains an open subset (in terms of suitable Sobolev norms) of data which develop into solutions that are conformally well behaved at future and past time-like infinity 
 shows that the existence of smooth conformal boundaries can be a fairly general feature of solutions to Einstein's field equations. 
 Besides a smallness condition there are no restrictions on the conformal Weyl tensor.
 
Recovering from the technical struggles that led to this insight, I began to wonder: 

\vspace{.2cm}

\noindent
{\it If the field equations ensure a smooth future evolution of $\Omega$ and $g$
beyond ${\cal J}^+ = \{\Omega = 0\}$, so that they define in the future of  
${\cal J}^+$ another `physical' solution to Einstein's field equations 
with metric $\hat{g} = \Omega^{-2}\,g$,  
and any gravitational radiation, represented by non-linear perturbations of the conformal Weyl tensor, travels unimpeded across ${\cal J}^+$  into that domain, 
why should physics come to an end at the future conformal boundary ${\cal J}^+$ 
?

} 

\vspace{.2cm}

This behaviour may be considered as just another quirk of the field equations, that should not be taken too seriously. The history of General Relativity shows, however,  that Einstein's equations were often wiser than their solvers. Physicists
made sense of the more exotic features 
of the solutions found by Schwarzschild and Friedmann 
only years after their discovery. 

Though it can all be found  in the article referred to above,
I never explicitly speculated about this in public. 
Being a beginner, it would hardly have been taken seriously, in particular,  because I had no answers to the questions:

\vspace{.1cm}

\noindent
$-$ {\it What happens if there is matter around} ? 

\vspace{.1cm}

\noindent
$-$ {\it How will matter behave in the far future} ?

\vspace{.1cm}

\noindent
Our discussion of the FLRW solutions above have shown that different matter models, exemplified there by $\rho_*$ and $w_*$, may have quite diverse consequences. Moreover,

\vspace{.1cm}

\noindent
 $-$ {\it What could be the meaning of the solution `on the other side' of ${\cal J}^+$ }?

\vspace{.1cm}

\noindent
In the stability result above the space-time defined by the metric $\hat{g} = \Omega^{-2}\,g$
{\it on the other side of ${\cal J}^+$} looks like a time reversed version of the  space-time end `on this side': from being infinitely extended at 
 ${\cal J}^+$ its space sections begin to shrink. This is certainly quite different from our present idea about the beginning of a cosmological space-time.  
 So I kept returning to the first two questions over the following years. 

\vspace{.1cm}

The first stability result generalizing the one above
concerns the (in four space-time dimensions) conformally invariant 
 Maxwell- or Yang-Mills-equations \cite{friedrich:1991}. It shows:

\vspace{.1cm}

 \noindent
  {\it  The nonlinear vacuum stability result outlined above generalizes to the
coupled Einstein-$\lambda$-Maxwell-Yang-Mills equations. The perturbed solutions admit smooth conformal boundaries in the future and the past. The conformal field equations determine  smooth conformal extensions of the solutions beyond these boundaries}. 
 
 \vspace{.1cm}
  
 This result establishes a pattern for analyzing various other situations in which conformally covariant matter transport equations are coupled to the Einstein-$\lambda$ equation. 
 Christian L\"ubbe and Juan Valiente Kroon studied 
 the Einstein-$\lambda$-perfect-fluid equations with
the equation of state $\hat{p} = 1/3\,\hat{\rho}$ for   
pure (incoherent) radiation. These matter equations have in  common with the Maxwell equations that the energy-momentum tensor is trace-free and the conformal matter equations have the same form as the  `physical' version. They show \cite{luebbe:valiente-kroon:2013}: 
 
 \vspace{.1cm}

 \noindent
{\it The FLRW-solutions with the equations of state 
 of pure radiation and a smooth conformal boundary in the future
 are non-linearly future stable in the class of all 
 Einstein-$\lambda$-perfect-fluid solutions with this equation of state.
 The perturbed solutions admit a smooth conformal boundary in the future and a smooth conformal extension beyond.} 

 \vspace{.1cm}
 
A further example leading to a similar result is given by the Einstein equations coupled to the massless Vlasov matter equations \cite{Joudioux:Thaller:Valiente Kroon}.

\vspace{.2cm}

In the FLRW models given by (\ref{cn-solution}), (\ref{sn-cn-dn-solution}) and the generalizations discussed above, 
forward Scri's and backward Scri's as well as Big Bangs and Big Crunches stand back to back in the conformal extensions.
 Motivated by the observations above, results on
  Paul Tod's ideas about isotropic singularities \cite{RPAC Newman:1993},
 \cite{tod:1987},
 \cite{tod:2003}, \cite{tod:2007}, where the initial singularity is represented after a suitable conformal rescaling by a finite space-like set similar to a ${\cal J}^+$, and by his thoughts about the nature of entropy near the big bang, 
Roger Penrose proposed a cosmological model, referred to as {\it Conformal Cyclic Cosmology} (CCC)  \cite{penrose:2011}. It considers a chain of universes 
 where a given universe, say $U_n$, develops in its future a well defined ${\cal J}^+$, referred to now as {\it the crossover surface}, which is followed by another universe, $U_{n + 1}$,  for which the crossover surface represents  an isotropic singularity. 
 The solutions (\ref{cn-solution}), (\ref{sn-cn-dn-solution}) can be used to create such situations.
 
 \vspace{.2cm}

 This again gives rise to complicated questions. Strongly simplifying assumptions on $U_n$ and $U_{n + 1}$ may provide situations where some kind of identification or  glueing of the different ends lead to a picture as outlined above. It is not clear, however, that anything similar can be done with any degree of control if more general, conformally curved, solutions to Einstein's equations are considered. The precise nature of the transition from the infinite future of $U_n$ to the beginning of $U_{n + 1}$
 is unresolved so far. It  should be brokered by a mechanism which guarantees a unique extension without any interference from outside, but not necessarily preserving the  time reflection invariance of  hyperbolic equations. 
This requires a closer look at the equations and the matter models from both sides of the crossover surface,  generalizing perhaps of the work initiated by Alain Bachelot \cite{bachelot:2020}.

\section{Conformally non-covariant matter fields.}

Finding such a mechanism, if something like it exists at all, 
requires among other things a sufficiently general and deep understanding of the behaviour of matter and the field equations at future time-like infinity. 
A number of authors analysed the future asymptotic behaviour of solutions to the Einstein-$\lambda$-perfect fluid equations
with an homentropic flow, where the entropy is
constant in space and time and the equation of state can be given in the form $\hat{p} = w(\hat{\rho})$ with some suitable function $w$.
Often they assume  a linear equation of state
$\hat{p} = w_*\,\hat{\rho}$, $w_* = const.$, and the additional condition $0 < w_* < 1/3$, see
\cite{hadzic-speck-2015}, \cite{oliynyk:2016},
\cite{reula:1999},
\cite{rodnianski:speck:2013}, \cite{speck:2012}. 
 In the articles  \cite{liu:wei} and \cite{reula:1999} is studied 
the future stability of FLRW space-times for more general classes of equations of state $\hat{p} = w(\hat{\rho})$. 
All of this work was done in more conventional representations of the field equations in terms of which the questions of interest here
may indeed be difficult to analyse.
None of the authors above looked at these things in the way indicated above.

\vspace{.1cm}

In the following I tried to generalize the kind of analysis begun above, hoping to answer the following question:

 \vspace{.1cm}

\noindent
 {\it Can there be achieved $C^{\infty}$, $C^{k}$, $\ldots$, or at least in some sense uniquely extendible conformal structures at future time-like infinity for solutions to 
Einstein-$\lambda$-matter equations with
 conformally non-covariant matter field equations
 ?}

 \subsection{Scalar fields.}

There are two results pointing into that direction.
In \cite{ringstroem:2008}
H. Ringstr\"om studied the future stability of a very general class Einstein-non-linear scalar field systems
with a
scalar field equation of the form
\begin{equation}
\label{scalar-field-equ}
\hat{\nabla}_{\mu}\hat{\nabla}^{\mu}\phi - \left(m^2\,\phi
+ V'(\phi)\right) = 0,
\end{equation}
where $' = \partial/\partial \phi$, and an energy momentum tensor
\begin{equation}
\label{scalar-field-EM-tensor}
\hat{T}_{\mu\nu} = \hat{\nabla}_{\mu}\phi\,\hat{\nabla}_{\nu}\phi 
- \left(\frac{1}{2}(\hat{\nabla}_{\rho}\phi\,\hat{\nabla}^{\rho}\phi + m^2\,\phi^2) + V(\phi)\right)\hat{g}_{\mu\nu},
\end{equation}
with a potential of the form $V(\phi) = \phi^3\,\mu + \phi^4\,U(\phi)$ where $\mu$ is a constant and
$U$ a smooth real-valued function. 
 In \cite{friedrich:massive-fields} has been 
considered a special case with the following result.

\vspace{.2cm}

\noindent
{\it If $\mu = 0$ and $3\,m^2 = 2\,\lambda$,
the coupled Einstein-$\lambda$-scalar-field equations
(\ref{einst}), (\ref{scalar-field-equ}), (\ref{scalar-field-EM-tensor})
 imply a reduced system of conformal field equations
  for the unknowns $\Omega$, $g_{\mu\nu} = \Omega^2\,\hat{g}_{\mu\nu}$,  $\psi = \Omega^{-1}\,\phi$ and some tensor fields derived from them. In a suitable gauge this is hyperbolic for any sign of 
  $\Omega$.
 Smooth Cauchy data for this system can be prescribed with $\Omega = 0$ on a compact $g$-space-like 3-manifold ${\cal J}^+$.
The development of these data backwards in time is smooth
and induces on a space-like slice $S$ (in the corresponding physical space-time) smooth standard Cauchy data $\Delta_0$
 for  the coupled Einstein-$\lambda$-scalar-field system. 
 For $\Delta_0$ there exist (in terms of Sobolev norms) an open neighbourhood of Cauchy data on $S$ for this system
 so that the future development of any smooth data in this neighbourhood admit a smooth conformal extension beyond the respective future time-like infinity.}

\vspace{.2cm}

We recall that the 
conformally covariant scalar operator 
$\phi \rightarrow L_{\hat{g}}\phi = (\Box_{\hat{g}} - 
\frac{1}{6}\,R[\hat{g}])\phi$
with $\Box_{\hat{g}} = \hat{\nabla}_{\mu}\hat{\nabla}^{\mu}$
satisfies
$L_{\Omega^2\hat{g}}(\Omega^{-1}\phi)
= \Omega^{-3}\,L_{\hat{g}}\phi$. 
With the trace of the energy momentum tensor  
(\ref{scalar-field-EM-tensor}) given by
\[
\hat{T} 
= - \hat{\nabla}_{\mu}\phi \,\hat{\nabla}^{\mu}\phi 
- 2\,m^2\,\phi^2 - 4\,V(\phi) = - R[\hat{g}] + 4\,\lambda,
\]
equation (\ref{scalar-field-equ}) can be expressed in terms of the conformally covariant wave operator. It reads then
\[
L_{\hat{g}}\phi = \left(m^2 - \frac{2}{3}\,\lambda\right)\phi   +
V'(\phi) - \frac{2}{3}\,\phi\,V(\phi)
- \frac{1}{6}\,\phi\,\hat{\nabla}_{\mu}\phi\,\hat{\nabla}^{\mu}\phi
- \frac{1}{3}\,m^2\,\phi^3,
\]
while we get in terms of the conformal fields
$\psi = \Omega^{-1}\,\phi$ and 
$g_{\mu\nu} = \Omega^2\,\hat{g}_{\mu\nu}$ the conformal representation
\[
L_{g}\psi =
\Omega^{-2}\left(m^2 - \frac{2}{3}\,\lambda\right)\psi  
+ \Omega^{-3}\,V'(\Omega\,\psi) 
- \frac{2}{3}\,\Omega^{-2}\,\psi\,V(\Omega\,\psi)
\quad \quad \quad \quad \quad \quad 
\quad \quad \quad \quad 
\]
\[
\quad \quad \quad \quad 
\quad \quad \quad \quad \quad \quad 
- \frac{1}{6}\,\psi\,
\nabla_{\mu}(\Omega\,\psi) \,\nabla^{\mu}(\Omega\,\psi) 
- \frac{1}{3}\,m^2\,\psi^3.
\]
Equation (\ref{scalar-field-equ}) is thus not conformally covariant 
with our conditions but {\it conformally regular} in the sense that there will remain no $\Omega^{-1}$ terms on the right hand side if the conditions of the theorem above are taken into account. This is a property it shares with the Einstein-$\lambda$ vacuum field equations.

The energy momentum tensor is not trace free but satisfies
\[
\hat{T} =
- \Omega^2\left(\nabla_{\mu}(\Omega\,\psi)\,\nabla^{\mu}(\Omega\,\psi) + \frac{4}{3}\lambda\,\psi^2 + 4\,\Omega^{- 2}\,V(\Omega\,\psi)\right)
\rightarrow 0 \,\,\,\mbox{as} \,\,\,\Omega \rightarrow 0,
\]
where the limit above will follow only {\it  if $g_{\mu\nu}$ and $\psi$ can be shown to extend smoothly as $\Omega \rightarrow 0$}. 
It should be mentioned that the hyperbolic reduced conformal field equations considered above as well as those considered in the following preserve the constraints implied by the conformal field equations. In the case of the present system, the proof of this fact is far from immediate.

\subsection{Pure dust cosmologies.}

The flow fields $\hat{U}$ of Einstein-$\lambda$-perfect-fluid solutions  with pure dust equation of state
\[
\hat{p} = 0,
\]
arising from data on a compact 3-manifold $S$ where $\hat{\rho} > 0$
are often interpreted as representing the cosmic flow of galaxies. 
The field equations imply that the trace of the corresponding energy momentum tensor satisfies  
\[
\hat{T} = - \hat{\rho} \neq 0.
\]
In the case of the conformally flat FLRW models we have seen that such solutions can develop smooth conformal boundaries. Whether this is also possible if the conformal Weyl tensor does not vanish is not  obvious (see  \cite{hadzic-speck-2015}).
In \cite{friedrich:dust:2016} it has been shown: 

\vspace{.2cm}

\noindent
{\it Consider a FLRW solution to the Einstein-$\lambda$-perfect-fluid  equation with pure dust equation of state arising from data 
on a Cauchy hypersurface $S \sim \mathbb{S}^3$.
If it admits a smooth conformal boundary at future time-like infinity, then any smooth general set of Cauchy data on $S$ for the same equation which is sufficiently close (in terms of Sobolev norms) to the FLRW-data develops into a solution that  is  time-like and null geodesically future complete, admits a smooth conformal boundary in its infinite future, where $\Omega  \rightarrow 0$, and extends 
as a smooth solution to the conformal field equations into a domain where 
$\Omega < 0$}. 

\vspace{.2cm}

The unknowns $g_{\mu\nu} = \Omega^2\,\hat{g}_{\mu\nu}$ and $\rho = \Omega^{-3}\,\hat{\rho}$ in the conformal field equations
then remain bounded as $\Omega  \rightarrow 0$ and it follows that 
\[
\hat{T} = - \Omega^{3}\rho \rightarrow 0
\quad \mbox{as } \quad \Omega  \rightarrow 0.
\]
Assuming initial data so that $\hat{\rho} > 0$,
the equation 
for the flow vector field  $\hat{U}^{\mu}$ reduces in the case of pure dust to the equation
\[
\hat{U}^{\mu}\,\hat{\nabla}_{\mu}\hat{U}^{\nu} = 0,
\]
and thus to an ODE. In terms of the conformal fields it takes the form
\[
0 = U^{\mu}\,\nabla_{\mu}\,U_{\nu}
- \Omega^{-1}\, 
(g^{\mu}\,_{\nu} + U^{\mu}\,U_{\nu})\nabla_{\mu}\Omega,
\]
and the $\Omega^{-1}$ term, which reflects the conformal non-covariance of the system, spreads into other equations of the conformal system.

\vspace{.1cm}

{\it The fact that the flow is geodesic combined with the particular structure of the energy momentum tensor of a perfect fluid allows us to make in this particular case contact with and exploit some conformally invariant structure}.

\vspace{.1cm}

Let $\hat{g}$, $\hat{U}$, $\hat{\rho} > 0$ satisfy the 
Einstein-$\lambda$-pure-dust equations so that 
$\hat{U}^{\mu}\,\hat{\nabla}_{\mu}\hat{U}^{\nu} = 0$.
If the functions $r$ and $q$ satisfy 
\[
0 \neq \hat{\nabla}_{\hat{U}} r = q\,r, \quad \quad 
0 \neq \hat{\nabla}_{\hat{U}}q = - \frac{1}{2}\,q^2 
+ (\lambda/6 - \hat{\rho}/3)\,r,
\]
the curve $x(s)$ with $\frac{d}{ds}x = V = r\,\hat{U}$ and the 1-form
$b = \hat{g}(q\,\hat{U}, \, \cdot \,)$ satisfy
the {\it conformal geodesic equations} 
\[\hat{\nabla}_VV + 2\,V<b, V> 
- \,\hat{g}(V, V)\,b^{\#} = 0,
\]
\[
\hat{\nabla}_Vb \,- <b, V>b + \frac{1}{2}\,
\hat{g}(V, \,\cdot\,)\,\hat{g}^{\#}(b, b) = \hat{L}(V, \,\cdot\,),
\]
where $\hat{L}$ denotes the Schouten-tensor of $\hat{g}$. 
The point of this observation is that the conformal geodesic equations are conformally invariant and the conformal geodesics are invariants of the conformal structure \cite{friedrich:cg on vac}.
This was used in \cite{friedrich:dust:2016} to regularize the conformal  Einstein-$\lambda$-pure-dust system near future time-like infinity.

The pure dust model is thus still {\it conformally privileged}.

\subsection{Equations of state with prescribed asymptotic behaviour.}

In the articles cited above that study the future behaviour of cosmological solutions to the Einstein-$\lambda$-perfect fluid equations on the basis of linear equations of state, $\hat{p} = w_*\,\hat{\rho}$, $w_* = const$, no arguments are given for this choice.
It appears to be rather a matter of convenience instead of being motivated by a deep understanding of its physical role.
While solutions to the Einstein-$\lambda$-perfect-fluid equations may provide good cosmological models on large scales,
it seems fairly unlikely that linear equations of state represent  natural 
requisites from the Big Bang to future time-like infinity. Models of the universe that behave at late times like a pure dust or a pure radiation FLRW model  or one of their conformally curved generalizations 
 may require transitions of the form
 \[
 \hat{p} = w^{**}(\hat{\rho})\,\hat{\rho},
 \]
with a function $ w^{**}(\hat{\rho})$ that satisfy, consistent with our earlier requirements,  $0 \le  w^{**}(\hat{\rho}) \le 1/3$ and assumes the value $w^{**}(\hat{\rho}) = 0$ or  $w^{**}(\hat{\rho}) = 1/3$
at late times. 
There is nothing, however,  which would fix a notion of  `late time'. The only meaningful requirement would be that these values are approximated in the limit when the space-time approaches future time-like infinity. 
The equation of state would then still need to recognize, however,
where and when this limit will be achieved.

\vspace{.2cm}

In the cases $(iii)$ of the FLRW models discussed above the 
physical density $\hat{\rho}$ and the conformal density $\rho$ satisfy a relation of the form
\begin{equation}
\label{rho-e-hat(rho)}
\hat{\rho} = \Omega^{e}\rho.
\end{equation}
In those cases we had $\rho = \rho_* = const. > 0$,
$e = const. > 0$ whence $\hat{\rho} \rightarrow 0$ as
$\Omega \rightarrow 0$ and $\hat{\rho} \rightarrow \infty$ as
$\Omega \rightarrow \infty$. The behaviour of $\hat{\rho}$ can thus be understood as an indicator for the approach to the infinite future or the Big Bang. In generalizing the situation we shall keep 
(\ref{rho-e-hat(rho)}), hoping that it will serve the indicator function at least in the far future where $\Omega \rightarrow 0$, but we may have to give up the relation $\rho = \rho_* = const.$
We could think of generalizing (\ref{rho-e-hat(rho)}) 
by assuming $e$ to be a function. In this article we will only consider the situations near the end where $\Omega \rightarrow 0$
and try to keep (\ref{rho-e-hat(rho)})
with $e = const. > 0$. This will work well as long as $\rho$ can  be guaranteed to stay positive and bounded as  $\Omega \rightarrow 0$. 

\vspace{.1cm}

The pure dust and the pure radiation equations of state are now generalized as follows.
An {\it asymptotic dust equation of state} is given by a function of the form
\begin{equation}
\label{as-dust-eos}
\hat{p} = w(\hat{\rho}) = \left(\hat{\rho}^{k}\,w^*(\hat{\rho})\right)
 \hat{\rho}\,
\quad \mbox{with some} \quad k \in \mathbb{N},
\end{equation}
combined with  (\ref{rho-e-hat(rho)}) where $e = 3$. It implies
with the notation $ ' = \partial/\partial \hat{\rho}$ 
\begin{equation}
\label{w'-as-dust}
w'(\hat{\rho}) = \hat{\rho}^{k}\,\left\{(1 + k) \,w^*
+ \hat{\rho}\,\,(w^*)'\right\}. 
\end{equation}

\vspace{.1cm}

\noindent
An {\it asymptotic radiation equation of state} is given by a function of the form 
\begin{equation}
\label{as-rad-eos}
\hat{p} = w(\hat{\rho}) = \left(\frac{1}{3} - \hat{\rho}^{k}\,\,w^*(\hat{\rho})
\right)\hat{\rho}
\quad \mbox{with some} \quad k \in \mathbb{N},
\end{equation}
combined with  (\ref{rho-e-hat(rho)}) where $e = 4$. It implies
\begin{equation}
\label{w'-as-rad}
w'(\hat{\rho}) = \frac{1}{3} - \hat{\rho}^{k}\left\{ 
(1 + k)\,w^*
+ \hat{\rho}\,(w^*)'\right\}.
\end{equation}
In both case $w^*(\hat{\rho})$ is assumed to be a smooth function defined  for all values of $\hat{\rho}$ that satisfies
\[
0  < w^*(\hat{\rho}) <  c = const.
\]
The positivity is required to clearly distinguish the asymptotic from the pure cases. Limits $w^*(\hat{\rho}) \rightarrow 0$ give back the pure dust and the pure radiation equations of state.

\vspace{.1cm}

The factors $\hat{\rho}^k$ with positive $k$ have been included in the definitions 
as a simple means to control the speed at which the pure dust or the pure radiation situations is approximated as $\hat{\rho} \rightarrow 0$.

\vspace{.1cm}

The conditions on $w^*$ may appear crude but they suffice for analysing the effect of the intended modifications of the equations of state in domains where 
$\hat{\rho}$ becomes small.
For $\hat{\rho}$ positive but sufficiently close to zero,
the range we are interested in, the terms in curly brackets
in (\ref{w'-as-dust}) and (\ref{w'-as-rad}) are positive. It follows that in the case of asymptotic dust the speed of sound $w'(\hat{\rho})$  is positive as 
$\hat{\rho} > 0$ and $w'(\hat{\rho}) \rightarrow 0$  as $\hat{\rho} \rightarrow 0$.
In the case of asymptotic radiation holds $w'(\hat{\rho}) > 0$ for 
$\hat{\rho} \ge 0$ and $w'(\hat{\rho})$ will remain positive if the solution can be smoothly extended into a domain where $\Omega < 0$.

\vspace{.1cm}

We note that {\it the Cauchy problems local in time for Einstein-$\lambda$-perfect fluids
with asymptotic dust or radiation equations of state pose no problems where $\hat{\rho}$ is sufficiently small}. This  follows from the results of
\cite{friedrich:1998}, \cite{friedrich:rendall} where only weak conditions on the equation of state are assumed.

\vspace{.1cm}

The principal parts  of the matter equations are affected by the equations of state above with the consequence that {\it any conformal covariance or privilege is lost}.
Definition (\ref{as-dust-eos})  implies 
 \[
\hat{T} = \hat{g}^{\mu\nu}\,\hat{T}_{\mu\nu} = 3\,w(\hat{\rho}) - \hat{\rho}
= 
- \Omega^{3}\,\rho + 3\,\Omega^{3 + 3\,k}\,\rho^{1+ k}\,w^*(\Omega^{3}\,\rho),
\]
while definition (\ref{as-rad-eos}) gives
\[
\hat{T} =  - 3\,\Omega^{\,4 + 4\,k}\,\rho^{1 + k}\,w^{*}(\Omega^{4}\,\rho).
\]
In both case $\hat{T} \neq 0$ if $\rho > 0$ and  
$\hat{T} \rightarrow 0$ as $\Omega \rightarrow 0$  if $\rho$ {\it remains bounded in this limit. } As seen below, this last condition will be met in the case of an asymptotic radiation equation of state while it is not clear  whether this can be guaranteed also in the case of an asymptotic dust equation of state.

\vspace{.1cm}

The conditions on the admissible values of $k$ required above can be weakened if the equations of state above are considered in the
conformal analogues of the Friedmann and energy conservation equation. In the case of asymptotic dust the system reads  
\[
(\dot{\Omega})^2 = \frac{\lambda}{3}
- \frac{R[\hat{h}]}{6}\,\Omega^2 + 
\Omega^{3}\,\frac{\rho}{3},
\quad \quad
\dot{\rho} = 3\,\Omega^{3\,k - 1}\,\rho^{1 + k}\,w^*(\Omega^3\,\rho))\,\dot{\Omega},
\]
in the case of asymptotic radiation the system reads 
\[
(\dot{\Omega})^2 = \frac{\lambda}{3}
- \frac{R[\hat{h}]}{6}\,\Omega^2 + 
\Omega^{4}\,\frac{\rho}{3},
\quad \quad 
\dot{\rho} = - 3\,\Omega^{4\,k-1}\,\rho^{1 + k}\,w^*(\Omega^4\,\rho))\,\dot{\Omega}.
\]
For suitable $k$ and initial data these equations can be smoothly
integrated across $\Omega = 0$ with $\dot{\Omega} < 0$ and 
$\rho$ bounded and positive. We shall see now that the FLRW assumption gives quite a wrong impression about the similarity and simplicity of these two cases if the assumption is dropped.

\subsubsection{On the equations with an asymptotic dust equation of state.}

To illustrate the kind of difficulties which arise in the case of 
an asymptotic dust equation of state we consider one of the equations 
contained in the complete system of conformal field equations.
If the latter are written in terms of an orthonormal frame $e_k$
with $e_0 = U$, the connection coefficient
$f_a = U^{\mu}\,e^{\nu}\,_a\,\nabla_{\mu}U_{\nu}$ is subject to the evolution equation
\[
e_0(f_a)  - w'\,e_a(\chi_c\,^c) + \chi_{ac}\,f^c + (1 - 3\,w')\,L_{a0} 
- w'\,\chi_c\,^c\,f_a 
\]
\[
= (1 - 3\,w')\,(\Omega^{-1}\,\nabla_0\Omega\,f_a 
+ \Omega^{-1}\,\chi_{ab}\,\nabla^b\Omega  
- \Omega^{-2}\,\nabla_0\Omega\,\nabla_a\Omega)
\]
\[
- w''\,\frac{\hat{\rho} + w}{w'}\,(\chi_c\,^c\,f_a 
- 3\,\Omega^{-1}\,\nabla_0\,\Omega\,f_a 
- \Omega^{-1}\,\chi_c\,^c\,\nabla_a\Omega
+ \Omega^{-2}\,\nabla_0\Omega\,\nabla_a\Omega),
\]
where $\Omega$, the connection coefficients $\chi_{ab} = \nabla_aU_b$ and $\chi_c\,^c = \nabla_kU^k$, and the Schouten tensor $L_{ik}$
obey further evolution equations.
In the case of pure dust the 
third line is not present. 
 In that case $w' = 0$ and the second line contains factors $\Omega^{-1}$. Even if the third
line is ignored, the arguments indicated above in the case of pure dust would not apply in the present case because the flow equation remains a partial differential equation if $w^* > 0$, $\hat{\rho} > 0$. It cannot be related to the conformal geodesics equations. 
The most complicated term in the equation is, however, the first factor in the third line. This is not even defined in the case of pure dust. The $\Omega^{-1}$ terms can not be compensated in this equation by suitable choices of $k$. We leave this case open.

\subsubsection{Solutions with an asymptotic radiation equation of state.}

In the case of an asymptotic radiation equation of state the situation is quite different. Consider the equation above again. All the coefficients introduced by the radiation equation of state are well defined and approach in the limit $w^* \rightarrow 0$ the corresponding value in the case of pure radiation. Moreover, the terms in the equation above that contain factors $\Omega^{-1}$ come with the coefficients 
 \[
1 - 3\,w' = 3\,(\Omega^{4}\, \rho)^k\left( (k + 1)\,w^* 
+ \Omega^4\, \rho\,(w^*)'\right) \quad \mbox{and}
\]
\[
w''\,\frac{\hat{\rho} + w}{w'} = - (\Omega^4\,\rho)^k \left(4 - 3\,\hat{\rho}^k\,w^*\right)
\frac{k\,(1 + k)\,w^* + 2\,(1 + k)\,\hat{\rho}\,(w^*)' 
+ \hat{\rho}^2\,(w^*)''}{
1 - 3\,(1 + k)\,\hat{\rho}^k\,w^* + 3\,\hat{\rho}^{1 + k}\,(w^*)'}.
\]
By a suitable choice of $k$ the factors $\Omega^{4\,k}$
thus allow us to make up for any negative power of $\Omega$.

The situation is similar with the other equations of the conformal system. Moreover, the way the 
asymptotic radiation equation of state affects the principal part of the system is so that one can still extract from the complete system in a suitable gauge a reduced system
that is symmetric hyperbolic irrespective of the sign of $\Omega$. 

\vspace{.3cm}

Solutions that admit smooth conformal extensions at future time-like infinity can now be constructed from data for the conformal field equations which are given on a space-like hypersurface $S$ in the `physical domain', where $\Omega > 0$ , or from data on the
hypersurface $S = \{\Omega = 0\} $ that represents future time-like infinity. To check that the data satisfy the constraints induced on
$S$, and possibly the additional requirements implied by the assumption that $\Omega = 0$ on $S$, these conditions are best expressed in terms of the unit normal to  $S$ and
then transformed into a frame $e_k$ with $e_0 = U$, whereby the evolution equations need to be used as well.
Unless the field $U$ is assumed to be orthogonal to $S$
this involves some fairly tedious calculations (see \cite{friedrich:nagy}, where the presence of a boundary requires them). Since the latter give limited insight they are skipped here. After the Cauchy problem has been solved for the given data it follows by
standard arguments that the constraints and thus the complete system of conformal field equations will be solved as well 
(see \cite{friedrich:dust:2016},  \cite{friedrich:rendall} for detailed discussions). We can state now the following results \cite{friedrich:2023:asymptotic-dust-radiation}.

\vspace{.2cm}

{\it The Einstein-$\lambda$-perfect-fluid equations with an asymptotic radiation equation of state where $k \ge 1$
induce  in a suitable gauge a  reduced system of the conformal  Einstein-$\lambda$-perfect-fluid equations that is symmetric hyperbolic irrespective of the sign of $\Omega$. 

\vspace{.1cm}

On a compact 3-dimensional manifold 
${\cal J}$ one can construct smooth Cauchy data for the reduced conformal equations with 
$\Omega = 0$, $U$ time-like future directed orthogonal to ${\cal J}$, 
and $<U, d\Omega> \,\,< 0$ that satisfy the constraints induced by the conformal field equations and the special requirements
on a space-like hypersurface on which $\Omega = 0$. 

\vspace{.1cm}

These data determine a smooth solution to the reduced equations with 
$U$ hypersurface orthogonal, 
$\Omega < 0$ in the future of 
${\cal J}$ and $\Omega > 0$ in the past of ${\cal J}$.
In the latter domain the solution defines a unique solution to the Einstein-$\lambda$-perfect-fluid equations with an asymptotic radiation equation of state 
that is time-like geodesically future complete and for which 
${\cal J}$ represents a conformal boundary at the infinite time-like future.

\vspace{.1cm}

Let $S$ be a Cauchy hypersurface for this solution in the past of 
${\cal J}$ and denote by $\Delta$ the Cauchy data induced by that solution on $S$. Any Cauchy data $\Delta'$ on $S$ for the same equations which are sufficiently close to $\Delta$ develop into a solution that is
also time-like geodesically future complete, admits a smooth conformal boundary in the future, and a smooth conformal extension beyond. }

\vspace{.1cm}

We note that the Cauchy hypersurface $S$ above is not required to be orthogonal to $U$ and that the flow vector field comprised by the data 
$\Delta'$ on $S$ is not required to satisfy the condition of hypersurface orthogonality.

\subsection{Concluding remarks.}

 If $k$ is large, the terms involving $w^*$
 in (\ref{as-dust-eos}) and (\ref{as-rad-eos}) may,  when $\Omega \rightarrow 0$, look 
like minor perturbations of the pure dust or pure radiation equations of state. We have seen, however, that 
the effects of these terms are quite different  in the two cases.

\vspace{.1cm}

 When $\Omega$ becomes small the term involving $w^*$ may
 in the case of the asymptotic radiation equations of state
indeed be considered already for $k = 1$ as a minor perturbation
 relative to the dominating first term on the right hand side of  (\ref{as-rad-eos}). This is apparently sufficient to preserve asymptotically the effects corresponding to the conformal covariance of the pure radiation equation of state.

\vspace{.1cm}

In contrast, the transition from the pure to the asymptotic dust equation of state represented by the  term involving $w^*$, comes with a drastic change of the principal part of the matter equations by which
the conformal privilege of the pure dust case is lost completely even if $k$ is large.

One could hope to simplify the analysis of this case by entangling the two problems that are possibly interfering in the asymptotic dust case at future time-like infinity.
Let the function $w^*$ be modified so that $w^*(\hat{\rho}) = 0$ precisely
if $\hat{\rho} = \Omega^3\,\rho$ falls below a certain positive threshold $\hat{\rho}_*$.
For the sake of discussion assume that the set 
$\{\hat{\rho} = \hat{\rho}_*\}$ defines a space-like Cauchy hypersurface
with the set  $\{\hat{\rho} > \hat{\rho}_*\}$, on which we have
an asymptotic dust equation of state, lying in its past and  
the set  $\{\hat{\rho} < \hat{\rho}_*\}$, on which we have
a pure dust equation of state, lying in its future.  
For this picture to make sense one has to decide whether the space-time evolution extends with a sufficient degree of smoothness across the set
$\{\hat{\rho} = \hat{\rho}_*\}$ where the change of principal part takes place which reduces the PDE for the flow field to an ODE.
Since this concerns the physical domain, one could expect the answer to follow from the analysis of the fluid equations in
\cite{friedrich:1998}, \cite{friedrich:rendall} which treats the case $w > 0$ and $w = 0$ along similar lines. If the
answer is positive, the problem of asymptotic smoothness at time-like infinity only concerns the fields on $\{\hat{\rho} < \hat{\rho}_*\}$. This is the situation considered in \cite{friedrich:dust:2016}. 
The question of interest now is 
whether the analyses at the set $\{\hat{\rho} = \hat{\rho}_*\}$ 
and of the behaviour at time-like infinity can be combined
to clarify what happens if the value $\hat{\rho}_*$ of the threshold  is lowered to eventually perform the limit 
$\hat{\rho}_* \rightarrow 0$.

In the light of the preceding discussions I consider this question   
as particular interesting. While it may not inform us about the final nature  of the matter fields at the end of our present universe it 
may give rise to a more subtle and appropriate definition of an asymptotic dust equation of state and 
will certainly give insights
into the freedom allowed by the field equations and the matter equations to model possible ends at future time-like infinity.

%\vspace{1cm}

}

\end{document}